\documentclass{article}
\usepackage{arxtools}

\begin{document}

\ArticleTitle
  {Optical states with higher stellar rank}
\ArticleAuthor*
  [0000-0002-5646-6964]
  {Jan Provazn\'{i}k}
  {provaznik@optics.upol.cz}
\ArticleAuthorAddress
  {Department of Optics, Palack\'{y} University, 17. listopadu 1192/12, 771 46 Olomouc, Czech Republic}

\ArticleAuthor*
  [0009-0001-3898-440X]
  {Olga Solodovnikova}
  {olgasol@dtu.dk}
\ArticleAuthorAddress
  {Center for Macroscopic Quantum States (bigQ), Department of Physics, Technical University of Denmark, Building 307, Fysikvej, 2800 Kgs. Lyngby, Denmark}

\ArticleAuthor*
  [0000-0003-4114-6068]
  {Radim Filip}
  {filip@optics.upol.cz}
\ArticleAuthorAddress
  {Department of Optics, Palack\'{y} University, 17. listopadu 1192/12, 771 46 Olomouc, Czech Republic}

\ArticleAuthor*
  [0000-0002-5761-8966]
  {Petr Marek}
  {marek@optics.upol.cz}
\ArticleAuthorAddress
  {Department of Optics, Palack\'{y} University, 17. listopadu 1192/12, 771 46 Olomouc, Czech Republic}

\ArticleTitlePrint

\begin{abstract}\noindent
  Quantum non-Gaussian states of traveling light fields are crucial components of quantum information processing protocols; however, their preparation is experimentally challenging. In this paper, we discuss the minimal requirements imposed on the quantum efficiency of photon number resolving detectors and the quality of the squeezing operation in an experimental realization of certifiable quantum non-Gaussian states of individual photonic states with three, four, and five photons.
\end{abstract}


%

\section{Introduction}

Quantum non-Gaussian states of traveling light fields are the lifeblood of scalable fault-tolerant quantum computation protocols employing continuous-variable quantum systems~\cite{lloyd1999,gottesman2001,menicucci2014,baragiola2019,bourassa2021,madsen2022,aghaeerad2025}. Unfortunately the production of these states is experimentally challenging due to the scarcity of naturally occurring non-Gaussian interactions. The lack of physical non-Gaussian interactions can be resolved with measurement-induced operations~\cite{filip2005,marek2009,marek2011,yukawa2013b,miyata2016,marek2018a,sakaguchi2023}, which can be implemented using Gaussian interactions and appropriate ancillary non-Gaussian states. The bespoke non-Gaussian ancillaries can be tailored precisely to their purpose and synthesized with several methods, for example, by utilizing photon addition and subtraction~\cite{dakna1999,fiurasek2005,eaton2019,takase2021,endo2023}, suitable manipulation and partial measurement of entangled states~\cite{yukawa2013a,yoshikawa2018,tiedau2019,provaznik2020}, and exploiting Gaussian boson samplers~\cite{su2019,quesada2019}. Quantum non-Gaussian states with greater complexity can be also crafted from simpler states by employing intricate breeding protocols~\cite{weigand2018,eaton2022,zheng2023,takase2024,aghaeerad2025}.

The essential non-Gaussian states can be expressed as finite superpositions of photon number states, possibly modified by Gaussian operations~\cite{lachman2019,fiurasek2022,walschaers2021,chabaud2020}. Their quantum non-Gaussianity is always hierarchically bounded by their stellar rank, which corresponds to the highest photon number in the minimal superposition~\cite{walschaers2021,lachman2019,fiurasek2022}, which is, in turn, bounded by the number of photons detected during their generation. So far, experimental preparation of non-Gaussian states of traveling light has been limited to detecting at most four photons~\cite{engelkemeier2021,endo2025b}, mainly due to the exponentially decreasing probability of success.
However, only detecting the required number of photons is insufficient to reach the genuine quantum non-Gaussianity of the desired rank. The omnipresent loss in realistic experiments negatively affects the prepared states and can obliterate their critical properties. It is necessary to certify that the states possess the desired qualities to guarantee they are suitable for further applications. Often, this cannot be sensitively assessed using fidelity or Wigner negativity; fidelity makes sense only for states close to their ideal counterparts, whereas Wigner negativity can only reveal their non-Gaussian nature and cannot be used to certify their stellar rank~\cite{walschaers2021}. Even though this situation is far removed from the current experimental reality, it is possible to at least lower-bound the stellar rank by employing witnesses based on hierarchical quantum non-Gaussian criteria~\cite{lachman2019,fiurasek2022}.

Photon number states represent typical states of a specific stellar rank. Apart from their fundamental role in describing quantum harmonic oscillators, they serve as a resource in quantum information processing~\cite{marek2009}, quantum metrology applications~\cite{kunitski2019,oh2020}, and can be directly employed in preparation of optical coherent Schr\"{o}dinger and Gottesman-Kitaev-Preskill states~\cite{winnel2024}. Their preparation can be viewed as a stepping stone towards generating quantum states that enable universal fault-tolerant optical quantum computing. Their stellar rank is also fairly robust when faced with loss; ideal photon number states never lose stellar rank due to loss~\cite{lachman2019}. Even though the probability of the initial photon number decays, it necessarily remains observable longer for photon number states than in any other finite photonic superposition. This behavior makes photon number states prime candidates for the role of test subjects in the analysis of various quantum state preparation protocols.

In this paper, we present a framework for evaluating the feasibility of experimental preparation of photonic states with a guarantee of the respective stellar rank, as witnessed by observing genuine quantum non-Gaussian behavior of the respective rank~\cite{lachman2019}. We investigate the limits on tolerable optical loss in specific experimental scenarios that still permit a successful preparation of three-, four-, and five-photon number states with a sufficient probability of success.
 
This document comprises two primary sections:  methodology and discussion of the findings. The first section introduces a mathematical model of the measurement-based photonic circuit capable of preparing individual photonic states of light, discusses the hierarchical criteria used to certify genuine quantum non-Gaussian states, and describes the Monte Carlo simulation of the experiment and the subsequent interpretation of its results. Figures presented in the second section determine the minimal requirements for the efficiency of the optical components, such as squeezing and detection, in experimental realizations targeting states of three, four, and five photons.

%

\section{Heralding Fock states of travelling light}

\begin{figure}[h]
  \begin{center}
    \includegraphics[width = 1.00 \columnwidth]{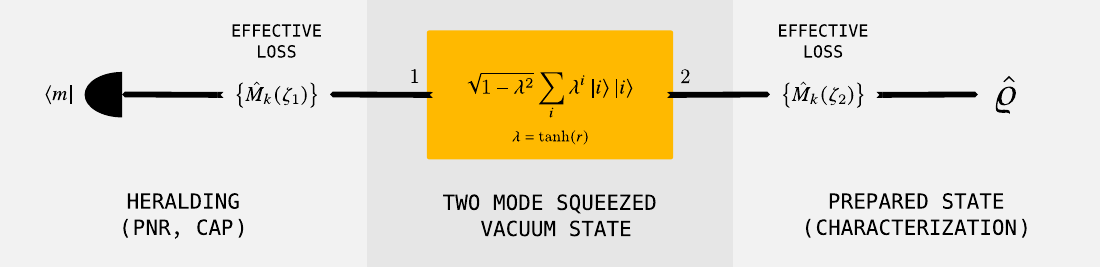}
  \end{center}
  \caption{
    Schematic illustration of the non-Gaussian state preparation circuit with a two mode squeezed vacuum state serving as a source of entangled quantum states. One of the modes is measured, thus projecting the other mode onto the resolved Fock state. The effective loss, characterized by the intensity transmittances $\zeta_{1}$ and $\zeta_{2}$, aggregate the various sources of loss present in generation of the entangled state, propagation and mode matching of its output modes, and detector efficiencies.
  }
  \label{f-scheme}
\end{figure}

Photonic number states can be conditionally prepared using a two-mode squeezed vacuum state and a photon number resolving detector~\cite{yukawa2013a,yoshikawa2018,tiedau2019,provaznik2020}. The theoretical model of the experimental scheme is illustrated in~\figref{f-scheme}. One of the entangled modes is measured, thus projecting the other mode onto the resolved Fock state. This procedure is repeated until a satisfactory detection outcome in the heralding mode is observed; at this point, the target state is successfully prepared. In the case of an unfavorable detection event, the state is discarded, and the whole procedure is repeated.

One of the principal experimental challenges in preparing quantum states of light is the omnipresent loss and noise. It diminishes quantum correlations between entangled resources and reduces the quantum efficiency of detectors, making it impossible to prepare the desired quantum state with perfect fidelity. The reduced detection efficiency negatively affects the preparation of the quantum state and its subsequent characterization. The procedure outlined in~\figref{f-scheme} assumes a perfectly squeezed state and an ideal photon number resolving (PNR) detector. The adverse effects of realistic inefficiencies can be accounted for by considering a lossy transmission of both modes prior to their measurement with ideal detectors~\cite{feito2009}. 

The resulting non-normalized marginal state is given by
\begin{equation}\label{e-pnr-rho}
  \hatrho_{(m)} \approx
  \sum_{i = 0}^{\infty} 
  \sum_{j = 0}^{\infty}
    \lambda^{i} \lambda^{j}
    \left(
      \sum_{k = 0}^{\infty}
        \bra{m} \hat{M}_{k} (\zeta_{1}) \ketbra{i}{j} \hat{M}_{k}^{\dagger} (\zeta_{1}) \ket{m}
    \right)
    \left(
      \sum_{k = 0}^{\infty}
        \hat{M}_{k}(\zeta_{2}) \ketbra{i}{j} \hat{M}_{k}^{\dagger} (\zeta_{2})
    \right) \Qc
\end{equation}
where~$m$ identifies the detected number state,~${\lambda = \tanh(r)}$ characterizes the two-mode pure squeezed state with initial squeezing~$r$, and the Kraus operators~${\hat{M}_{k} (\zeta)}$ describe the transmission loss with
\begin{equation}
  \hat{M}_{k} (\zeta) =
    \sqrt{ \frac{(1 - \zeta)^{k}}{k!} } 
    \sqrt{\zeta}^{\hat{n}} \hat{a}^{k}
\end{equation}
where~$\zeta$ gives the intensity transmittance of the lossy channel~\cite{ivan2011}. While loss is the primary obstacle in optical experiments, additional detrimental effects, such as thermal noise, can be similarly modeled with appropriate quantum channels \cite{ivan2011,gagatsos2017}.

The parameter~$\zeta_{1}$ inside relation~\eqref{e-pnr-rho} corresponds to the loss in the first mode, called the heralding mode, whereas~$\zeta_{2}$ describes the loss affecting the mode carrying the prepared state. The probability of successful preparation, that is, the probability of detecting~$m$ photons in the heralding mode, can be obtained analytically as
\begin{equation}\label{e-pnr-pro}
  P_{(m)} = (1 - \lambda^{2}) 
  \frac
    { (\lambda^{2} \zeta_{1})^{m} }
    { [ 1 - \lambda^{2} (1 - \zeta_{1}) ]^{m + 1} } \Qd
\end{equation}
The equation~\eqref{e-pnr-rho} for the resulting density matrix of the prepared state can be simplified. The matrix is diagonal in the Fock basis; its properly normalized elements are obtained as
\begin{equation}\label{e-pnr-rho-kk}
  \braket{ k | \hatrho_{(m)} | k } =
  \frac
    { [ 1 - \lambda^{2} (1 - \zeta_{1}) ]^{m + 1} }
    { [ \lambda^{2} (1 - \zeta_{1}) ]^{m} }
  \left( \frac{ \zeta_{2} }{ 1 - \zeta_{2} } \right)^{k}
  H (k, m, x) \Qc
\end{equation}
where the substitution~${x \coloneqq \lambda^{2} ( 1 - \zeta_{1} )(1 - \zeta_{2} )}$ is used in the function~$H(k, m, x)$ defined as
\begin{equation}\label{e-H}
  H(k, m, x) \coloneq
  \sum\limits_{l = \tau_{km}}^{\infty}
    \binom{l}{m}
    \binom{l}{k}
    x^{l} 
\end{equation}
with~${\tau_{km} \coloneqq \max(k, m)}$. The infinite series in~\eqref{e-H} is convergent as the substituted argument~$x$ satisfies~${\abs{x} < 1}$. It can be equivalently expressed using a suitable hypergeometric function~\cite{bateman1981}
\begin{equation}
  \begin{aligned}
    H(k, m, x) & =
    x^{\tau_{mk}} 
    \binom
      {\tau_{mk}}
      {\kappa_{mk}}
    {}_{2}F_{1} (
      1 + \tau_{mk},
      1 + \tau_{mk},
      1 + \tau_{mk} - \kappa_{mk},
      x
    ) \\
    & =
    x^{\tau_{mk}} F(k, m, x),
  \end{aligned}
\end{equation}
where~${\kappa_{mk} \coloneqq \min (m, k)}$. The hypergeometric function can be efficiently evaluated with the help of commonly available numerical libraries~\cite{virtanen2020}. 

The formula \eqref{e-pnr-rho-kk} deserves further discussion. The first two fractions seemingly diverge in the ideal lossless cases when either~${\zeta_{1} \to 1}$ or~${\zeta_{2} \to 1}$. It may also diverge if there is no initial two-mode squeezing applied, that is, when~${\lambda \to 0}$. The offending elements can be propagated into the~$H(k, m, x)$ function, leading to the full expression
\begin{equation}
  { [ 1 - \lambda^{2} (1 - \zeta_{1}) ]^{m + 1} }
  { \zeta_{2}^{k} }
  \left(
    \lambda^{2 (\tau_{mk} - m)}
    (1 - \zeta_{1})^{\tau_{mk} - m}
    (1 - \zeta_{2})^{\tau_{mk} - k}
  \right)
  F (k, m, x),
\end{equation}
where both~${\tau_{mk} - m \geq 0}$ and~${\tau_{mk} - k \geq 0}$ in the potentially diverging powers of the~${1 - \zeta_{1}}$, ${1 - \zeta_{2}}$ and~$\lambda^{2}$ coefficients.

\subsection*{Approximate photon number resolving detectors}

The model of state preparation, represented by relation~\eqref{e-pnr-rho}, relies on true PNR detectors in the heralding stage of the circuit. While these detectors exist in principle~\cite{hopker2019,endo2021,endo2025a}, their practical availability is severely limited. The model of the experimental circuit can be extended to cover the more commonly available cascaded avalanche photodiode (CAP) detectors~\cite{hlousek2019,grygar2022,hlousek2024,ercolano2024} that operate by dividing the incoming signal equally between~$n$ independent avalanche photodiode detectors and counting the number~$m$ of detectors that registered photons. The avalanche detectors can only distinguish between zero and non-zero numbers of incident photons; this hinders the possibility of photon number resolution by the cascade. It approximates true PNR detectors well only in the limit of large numbers~$n$ of the constituent avalanche detectors~\cite{provaznik2020}. The detection events when~$m$ out of the~$n$ detectors register photons are associated with the operators~\cite{paul1996}
\begin{equation}\label{e-cap}
  \hat{\Pi}_{m}^{n} 
    \coloneqq \sum_{i = m}^{\infty} w(i, m, n) \ketbra{i}{i} \Qc
\end{equation}
where the weights~$w(i, m, n)$ define the probabilities of~$i$ incident photons triggering any~$m$ constituent detectors out of the total number~$n$ of detectors making up the cascade, given by
\begin{equation}\label{e-cap-w}
  w(i, m, n) \coloneqq 
    \frac{1}{n^{i}} \binom{n}{m} \sum_{j = 0}^{m} \binom{m}{j} (-1)^{j} (m - j)^{i} 
    \quad\text{when}\quad i \geq m \Qd
\end{equation} 
The relations~\eqref{e-pnr-pro} and~\eqref{e-pnr-rho} can be readily adapted to CAP detectors. The resulting probability of success, obtained when heralding on the~$\hat{\Pi}_{m}^{n}$ outcome, 
\begin{equation}\label{e-cap-pro}
  P_{(m, n)} = \sum_{i = 0}^{\infty} w(i, m, n) P_{(i)} \Qc
\end{equation}
is the weighted average of the individual probabilities. The operator~\eqref{e-cap} of the measurement outcome and the original density matrix~\eqref{e-pnr-rho-kk} are diagonal in Fock representation; the resulting density matrix retains its diagonal structure. It is determined as a weighted average,
\begin{equation}\label{e-cap-rho-kk}
  \hatrho_{(m, n)} =
    \frac{1}{ P_{(m,n)} }
    \sum_{i = 0}^{\infty} w(i, m, n) P_{(i)} \hatrho_{(i)}
  \Qd
\end{equation}

%

\subsection*{Genuine quantum non-Gaussian states}

Realistic inefficiencies in all the stages of the state preparation procedure are modeled with lossy channels. The reduced heralding efficiency makes it impossible to create the desired state with perfect fidelity. Effects of loss incurred during its characterization further diminish the already imperfect state. This can be partly alleviated using hierarchical criteria of genuine quantum non-Gaussianity~\cite{lachman2019}, closely related to the stellar rank~\cite{chabaud2020,walschaers2021,fiurasek2022} of the prepared state, as their invariance under Gaussian operations offers a certain degree of robustness against loss~\cite{lachman2019}.

Genuine~$m$-photon quantum non-Gaussian ($m$-PQnG) states~\cite{lachman2019} cannot be decomposed into statistical mixtures of pure quantum states attainable by Gaussian transformations of superposed photon number states up to~${\ket{m - 1}}$. Whether a quantum state belongs to a particular~$m$-PQnG hierarchy can be determined by examining its first~$m$ photonic contributions. The original witness of quantum non-Gaussianity~\cite{lachman2019} can be equivalently expressed as constraint~\cite{fiurasek2022} imposed on a pair of computed quantities
\begin{equation}
  x_{m} 
    = \sum_{k = m + 1}^{\infty} 
      \braket{k | \hatrho | k}
    = 1 - \sum_{k = 0}^{m} 
      \braket{k | \hatrho | k}
  \quad\text{and}\quad
  y_{m} = \braket{m | \hatrho | m}
  \Qd
\end{equation}
If the quantities satisfy the inequality~${y_{m} \geq F_{m} (x_{m})}$, the density operator~$\hatrho$ represents a genuine~$m$-PQnG state. The threshold function~$F_{m}(x)$ in the inequality can be obtained through sophisticated numerical optimization~\cite{lachman2019,fiurasek2022}.

%

\subsection*{Monte Carlo simulation of the state preparation circuit}

The maximal tolerable amount of loss present in the optical circuit capable of preparing a certifiable genuine~$m$-PQnG state can be determined by numerically simulating the experiment. The measurement statistics of a realistic experiment can be emulated using a Monte Carlo simulation based on the theoretical model of the circuit. The relations~\eqref{e-pnr-rho-kk}~and~\eqref{e-cap-rho-kk} determine the diagonal elements of the conditionally prepared quantum state depending on the choice of heralding detectors used in the experiment. Both relations are functions of the initial squeezing rate~$r$, the transmittances~$\zeta_{1}$ and~$\zeta_{2}$ of the lossy channels hindering both modes, and the post-selection criteria imposed on the measurement outcome~$m$.

The choice of the heralding detectors does not affect the simulation methodology and the subsequent state certification. The following description is therefore given in terms of the more straightforward relation~\eqref{e-pnr-rho-kk} derived for true PNR detectors.

The preparation circuit is evaluated for different amounts of loss in both modes, determined by the transmittances~$\zeta_{1}$ and~$\zeta_{2}$ with varying initial squeezing rates~$r$ and distinct target states~${m}$. Detection events in the simulated experiment are drawn as random samples from a multinomial distribution bootstrapped with the diagonal elements~$\braket{k|\hatrho_{(m)}|k}$ (where~${0 \leq k < 20}$) of the computed density matrix~\eqref{e-pnr-rho}. 

The number of random samples reflects the probability of successful preparation~$P_{(m)}$. The probability~\eqref{e-pnr-pro} depends on the initial squeezing rate and the loss in the heralding mode. Given a budget of~$10^{8}$ repetitions available in a single realistic experimental run, only~${\lfloor 10^{8} \times P_{(m)} \rfloor}$ samples are drawn from the distribution and used to estimate the experimental probability distribution~${\overline{p}_{k} \approx \braket{k|\hatrho_{(m)}|k}}$. This random sampling process is repeated~$1000$ times to obtain an ensemble of independent experimental runs for the subsequent statistical analysis.

Certification of the simulated states is achieved with the hierarchical criteria~\cite{lachman2019}. The estimated experimental probabilities~$\overline{p}_{k}$, obtained in each simulated run of the experiment, are used to compute the aggregate random variables characterizing the experimental state,
\begin{equation}
  x_{m} = 1 - \sum_{k = 0}^{m} \overline{p}_{k} 
  \quad\text{and}\quad
  y_{m} = \overline{p}_{m} 
  \Qd
\end{equation}
Their expectation values and standard deviations, obtained from the ensemble of independent experimental runs, are then used to certify the quantum state resulting from the simulation with the particular choice of~${(m, \zeta_{1}, \zeta_{2}, r)}$ parameters. The experimental state is considered a certifiable genuine~$m$-PQnG state if the expectation values lie at least \emph{three standard deviations} from the threshold function~$F_{m} (x)$ in both axial directions.

\begin{figure}[h]
  \bgroup
    \hspace*{-0.125\columnwidth}%
    \includegraphics[width = 1.25 \columnwidth]{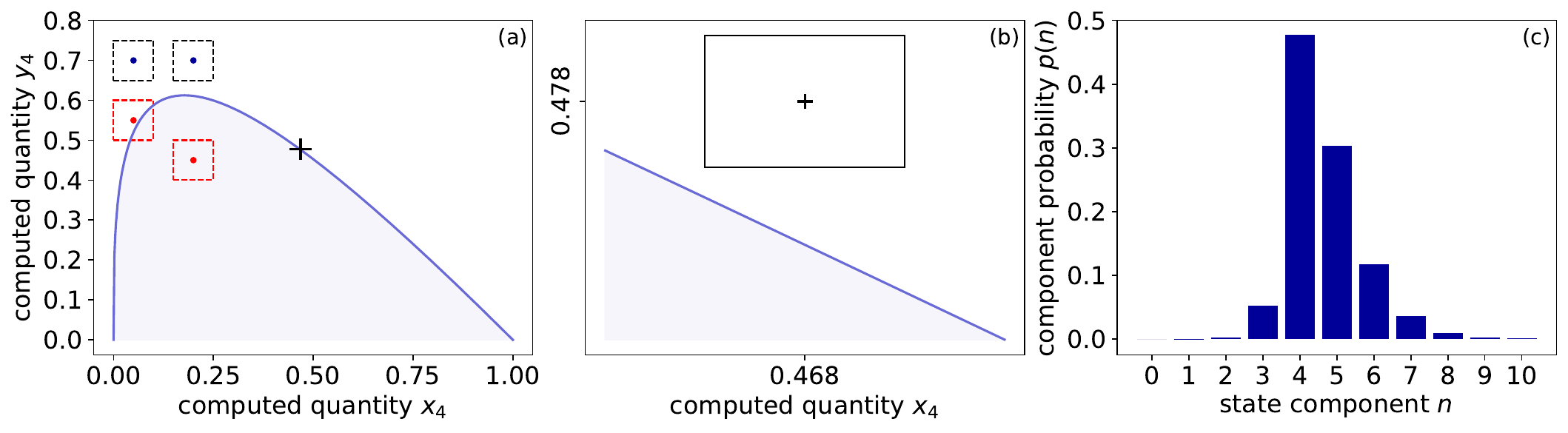}
  \egroup
  \caption{
    Illustration of the certification procedure of genuine~$4$-PQnG states for the simulated statistical ensembles. \textbf{(a)}~The blue line represents the threshold curve~$F_{4} (x)$. States with~$y_{4} > F_{4}(x_{4})$ are genuine~$4$-PQnG states. Four example experimental states are depicted in the figure. The bullet points represent the expectation values obtained by simulating the experiment. The dashed boxes represent their uncertainty and span three standard deviations in both axial directions. The pictured boxes are exaggerated in size for the legibility of the illustration. States marked with red bullets failed the certification as they lie either under the threshold curve or their respective uncertainty boxes intersect the curve. States marked with blue bullets are certifiably genuine~$4$-PQnG states according to the hierarchical criteria; their boxes are well above the curve and do not intersect the threshold curve. The black cross above the threshold curve marks a~$4$-PQnG state corresponding to~$2.75\%$ loss in characterization and~$20\%$ loss in heralding. Its certification is visualised in \textbf{(b)}~where the solid box represents the actual three sigma uncertainties in both computed quantities. \textbf{(c)}~The photon number distribution of the prepared quantum state. It is contaminated with higher-ordered contributions due to the loss incurred in heralding and lower-ordered components caused by characterization loss.
  }
  \label{f-il}
\end{figure}

The operating principle of the certification procedure is illustrated in~\figref{f-il}[a], with several example states shown, both passing and failing the certification, along with the actual threshold curve for genuine~$4$-PQnG states. States in the shaded area below the curve fail their certification.

The \figref{f-il}[b] depicts the certification of a simulated genuine~$4$-PQnG state, indicated by the black cross marker in \figref{f-il}[a]. The solid box in \figref{f-il}[b] represents the actual three sigma uncertainty in both computed quantities. Photon number distribution of the state is presented in \figref{f-il}[c]. The significant contributions of higher-ordered components are caused by~$20\%$ loss in the heralding phase of its preparation. Even though the contribution of four photons certainly majorizes all the other components, the quality of the state cannot be reliably evaluated using fidelity. Its value, just below~$50\%$, does not inspire confidence or support any concrete conclusion about the state. Yet, despite the substantial loss, it can be positively identified as a genuine~$4$-PQnG state with the hierarchical criteria. 

The extension of this methodology to include CAP detectors is mostly trivial. However, the infinite series in~\eqref{e-cap-pro} and~\eqref{e-cap-rho-kk} stemming from the expression~\eqref{e-cap} must be truncated accordingly. The number of elements considered in the series depends on the particular choice of the~$m$ and~$n$ parameters as these affect how the weight function~${w(i, m, n)}$ behaves. With~$m$ and~$n$ finite and close to each other, the weight function is generally heavy-tailed~\cite{provaznik2020}, and a sufficiently large cutoff for the summation has to be chosen.

%

\section{Discussion of results}

The main result of this work lies in finding out the maximal tolerable overall loss in the optical experiment capable of reliable production of certifiable genuine~$4$-PQnG and~$5$-PQnG states. The use of the hierarchical criteria~\cite{lachman2019} is a necessity due to the realistic loss in preparation of the two-mode squeezed states used in the experiment, in their propagation, and the low detection efficiency of contemporary detectors with photon number resolving capacity, amplified by the practical unavailability of true PNR detectors. These issues affect the prepared states negatively, manifesting in suboptimal fidelity with the desired state and leading to poor interpretability of the results. Fidelity alone cannot be reliably used to determine whether the prepared state is merely an attenuated state of four photons on something else entirely. Unlike fidelity, the hierarchical criteria, based on the stellar rank of the resulting state, offer concrete and unequivocal evidence of non-Gaussian behavior with a greater degree of robustness against loss~\cite{lachman2019}.

\begin{figure}[h]
  \bgroup
    \hspace*{-0.125\columnwidth}%
    \includegraphics[width = 1.25 \columnwidth]{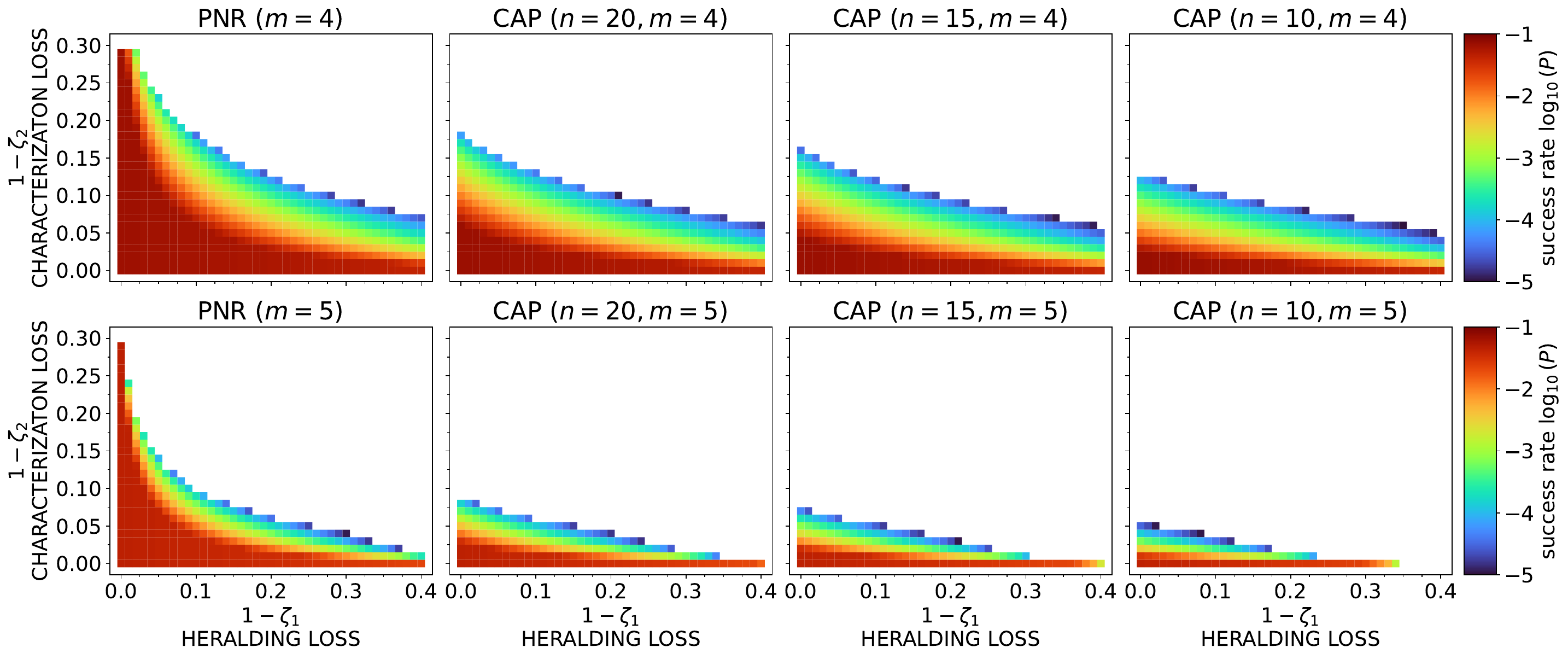}
  \egroup
  \caption{
    The tolerable loss in preparation of certifiable genuine~$4$-PQnG ({top}) and~$5$-PQnG ({bottom}) states. Individual tiles represent the best attainable probability of success. Values for each tile are obtained by maximizing the probability of successfully preparing a certifiable state over the initial squeezing rate~${0 \leq r \leq 10\ \dB}$. White colored tiles correspond to statistically insignificant cases with probabilities of success below~$10^{-5}$. Different heralding detectors were used in the analysis. The results obtained for a true PNR detector are presented in the leftmost column, while the remaining columns represent CAP detectors with~${n = 20, 15}$, and~$10$ constituent avalanche detectors.
  }
  \label{f-res-45}
\end{figure}

The results of the analysis of the experimental preparation of certifiable genuine~$4$-PQnG states are presented in the {top} row of \figref{f-res-45}. The analysis is done by considering different combinations of loss incurred during state preparation and its subsequent characterization. Certification of the prepared states accounts for realistic statistical behavior. The simulation reflects realistic experimental repetition rates. The cases with low probabilities of success (lower than~$10^{-5}$) leading to insufficient sample sizes (below 1000) are excluded from the presented results. Values in each tile are obtained by maximizing the probability of successfully preparing a certifiable state with respect to the initial squeezing rate~${0 \leq r \leq 10\ \dB}$.

Experimental preparation of genuine~$4$-PQnG states may just be feasible with state-of-the-art photon number resolving detectors~\cite{endo2021,endo2025a}. Using alternative CAP detectors further increases the requirements on the experiment; however, the simulations suggest that experiments using cascaded detectors composed of sufficiently large numbers of high-quality avalanche detectors are also within the realm of feasibility. Notably, while the differences in probability of success between CAP and PNR detectors are significant, the differences between CAP detectors with~${n = 10}$ and~${n = 20}$ are negiligible and mostly in the area of low heralding loss.

The analysis also extends to genuine~$5$-PQnG states in the {bottom} row of \figref{f-res-45}. The overall difficulty of the experimental realization increases significantly, rendering the realization using CAP detectors with lower numbers (${n = 10, 15}$) of constituent detectors essentially infeasible. Even when using as many as~${n = 20}$ detectors, the feasible region is extremely narrow, allowing for at most~$5\%$ of overall loss in its characterization while permitting non-zero loss during the preparation of the state.

This trend follows with states of higher order; this is only natural, as even in the ideal lossless case with a true PNR detector, the probability of success~\eqref{e-pnr-pro} scales with~${\lambda^{2m}}$ (where~${0 \leq \lambda < 1}$). Increasing the repetition rate in the analysis would balance the odds, as would increasing the number of detectors comprising the CAP detectors or using a true PNR detector.

\begin{figure}[h]
  \begin{center}
    \includegraphics[width = \columnwidth]{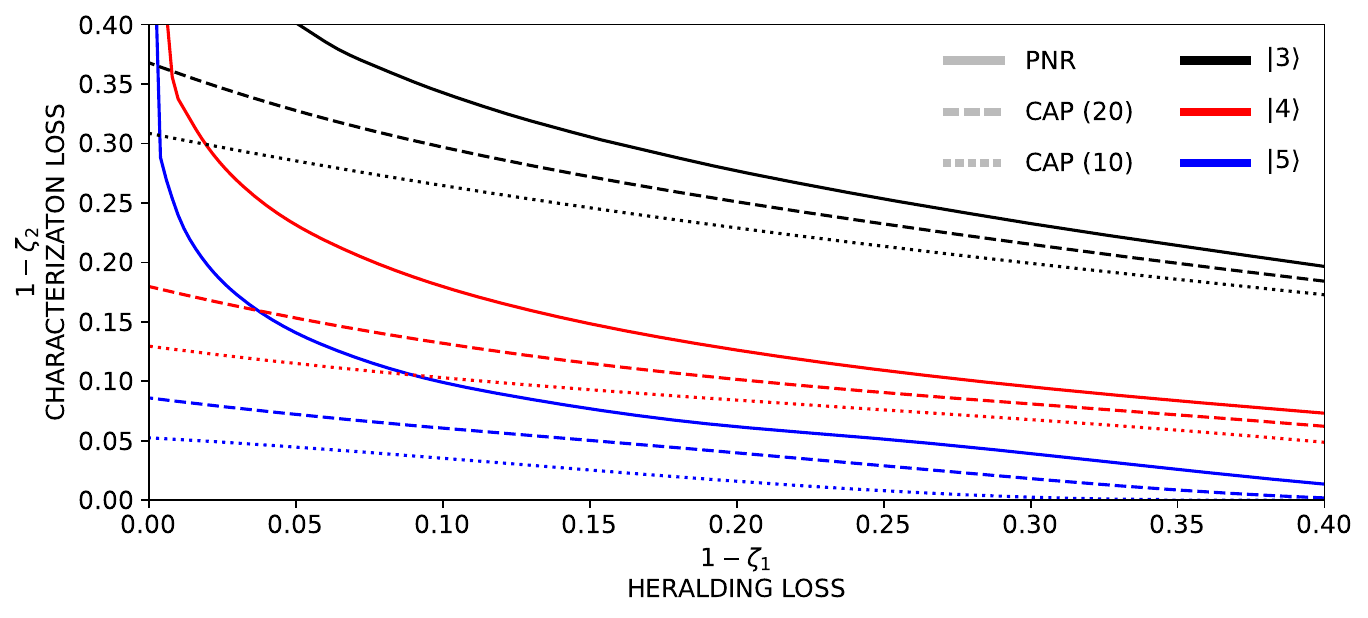}
  \end{center}
  \caption{
    Thresholds of the tolerable loss in preparation of certifiable genuine $3$~(black), $4$~(red), and $5$-PQnG (blue) states. Their values are determined using the same methodology as the results visualised in \figref{f-res-45} where the minimal viable probability of success is limited to~$10^{-5}$. Solid lines represent heralding with true PNR detectors, whereas the dashed and dotted lines correspond to CAP detectors comprising~$20$ and~$10$ photodiodes.
  }
  \label{f-thr-345}
\end{figure}

The colorful tiles presented in \figref{f-res-45} show the maximal probability of successful preparation of the certifiable genuine~$m$-PQnG states given the particular loss in heralding and characterization. \figref{f-thr-345} supplements this detailed analysis and provides a clear overview of the maximal tolerable loss by showing the actual thresholds between the feasible and infeasible regions, delineated by the probability of successful preparation greater than~$10^{-5}$, 
Three distinct sets of thresholds are presented for certifiable genuine~$3$-PQnG, $4$-PQnG, and~$5$-PQnG states in black, red, and blue colors (in order). Different detection methods used for heralding are distinguished with unique line styles. PNR detectors use solid lines, while CAP detectors with~$20$ and~$10$ constituent photodiodes are shown using dashed and dotted lines.

This figure demonstrates the stark difference between the requirements imposed on the preparation of different~$m$-PQnG states. For example, when targeting the~$5$-PQnG state and using a simpler CAP detector ($10$~constituent photodiodes, dotted line), the maximal tolerable loss in characterization is lower than~$5\%$. It quickly reaches zero for roughly~$30\%$ loss in heralding. Conversely, the thresholds for the~$3$-PQnG state are much greater, with the tolerable characterization loss exceeding~$20\%$ even when the heralding loss reaches~$40\%$. In this sense, while the~$4$-PQnG state requires far stricter experimental precision in comparison, it still permits considerably looser control than is needed to prepare the~$5$-PQnG state.

Generally, the higher the desired stellar rank, the lower the tolerable loss and the greater the demands on detector quality. Figures \ref{f-res-45} and \ref{f-thr-345} reinforce the evergreen that true PNR detectors are essential to the experimental endeavor. While CAP detectors can partly subvert this requirement, their greater susceptibility to loss makes or breaks the ambitious preparation of four and five photonic states.

%

\FloatBarrier
\section{Conclusions and outlooks}

The ever-present loss, combined with the limited efficiency of quantum detectors, is an unavoidable part of quantum-optical experiments affecting both practical and theoretical applications. The much sought-after quantum non-Gaussian states of traveling light, including those used in quantum computation~\cite{lloyd1999,gottesman2001,menicucci2014,baragiola2019,bourassa2021,madsen2022,aghaeerad2025}, quantum non-Gaussian state engineering~\cite{filip2005,marek2009,marek2011,yukawa2013b,miyata2016,marek2018a,sakaguchi2023} and quantum metrology~\cite{kunitski2019,oh2020}, require precise experimental control with as little loss and noise as possible.

In this work, photon number states are primarily employed as indicators of the quality of the exerted experimental control. Their use is motivated by their greater resilience against loss than the more complex quantum superpositions. The prepared states are impacted by a loss at every point of the process; higher-ordered components are inadvertently introduced by inefficient heralding detectors utilized in their preparation, whereas lower-ordered contributions appear due to inefficiencies in their characterization. Special hierarchical criteria of genuine quantum non-Gaussianity, based on stellar rank and offering innate resilience against loss, were employed to certify the prepared states. These criteria ascertain that the studied states cannot be produced with Gaussian operations from states with lower stellar rank. 

We demonstrated the maximal tolerable loss in the experimental realization of certifiable genuine~$4$ and~$5$-photon quantum non-Gaussian states, which, to the best of our knowledge, have yet to be experimentally prepared with traveling light fields. The analysis focused on true PNR detectors and their approximate versions, the more experimentally accessible CAP detectors. The promising results might serve as encouragement and perhaps as a basic guideline to our fellow experimental practitioners.
The analysis of tolerable loss in the presented scheme could be extended to states with higher stellar rank. Additionally, the hierarchical criteria of genuine quantum non-Gaussianity could be used to analyze other state preparation protocols and determine their loss tolerance.

%

\FloatBarrier
\section*{Data availability}

The software implementation of the numerical simulations is available through a public repository~\cite{source}. The repository also includes the software required to produce the visual representation of the results. The precomputed datasets necessary to exactly reproduce the figures presented within the manuscript are available through the same repository.

\FloatBarrier
\section*{Author contributions}
RF conceived the direction of the presented research. JP derived the semi-analytical model, conducted the numerical calculations and simulations, and performed the subsequent data processing and its visualisation. OS independently checked the correctness of the semi-analytical model using a distinct numerical methodology based on non-convex linear combinations of Gaussian states. All authors participated in preparation of the manuscript.

\FloatBarrier
\section*{Acknowledgements}

PM and JP acknowledge Grant No. 22-08772S of the Czech Science Foundation.
OS acknowledges funding from the Danish National Research Foundation (bigQ~DNRF~142).
RF, PM, JP, and OS acknowledge the European Union's HORIZON Research and Innovation Actions under Grant Agreement no. 101080173 (CLUSTEC).
RF, PM, and JP further acknowledge the Grant Agreements no. 101017733 and 731473 (CLUSSTAR).
RF and PM also acknowledge the project \mbox{CZ.02.01.01/00/22\_008/0004649} (QUEENTEC) of EU and MEYS Czech Republic. 
The authors acknowledge early discussions with Luk\'{a}\v{s} Lachman regarding the hierarchical criteria. The analysis initially employed previously unpublished threshold curves derived by Luk\'{a}\v{s} Lachman. 
OS thanks Jonas Schou Neergaard-Nielsen and Ulrik Lund Andersen for their insight and support. 
JP also acknowledges fruitful discussions with Jarom\'{i}r Fiur\'{a}\v{s}ek, Michal Matul\'{i}k, Vojt\v{e}ch Kala, and \v{S}imon Br\"{a}uer regarding witnesses of stellar rank. 
Additionally, JP acknowledges massive use of the computational cluster of the Department of Optics and using several open-source software libraries~\cite{hunter2007,harris2020,virtanen2020,dalcin2021} in the computation and subsequent evaluation of the presented results.

\FloatBarrier
\section*{Funding}

A list of funding sources is produced by the journal.

%

\FloatBarrier
\printbibliography

\end{document}